# Beta sheet propensity and the genetic code

Valentina Agoni

valentina.agoni@unipv.it

## Abstract

So far mutations analysis was performed in terms of transitions and trasversions, so on the basis of the molecule, or in terms of GC-content and isochors, through the quantification of GC→AT mutations over AT→GC mutations. We tried a different approach hypothesizing that probably we are partially protected (by the genetic code) from amyloid fibrils formation.

## The codon table

The analysis of mutation rate has been done in terms of transitions and trasversions or in terms of GC-content. Therefore the diverse approaches gave diverse results. Another possible explanation of the different percentages of the various mutations described in literature can be the fact that the mutations rate can be different in different site types (CpG, TpG or usual sites) (Mizawa 2012).

It is known that the amino acids coding table evolved to minimize the impact of deleterious mutations. In Agoni, 2013 we show that the organization of amino acids encoded by two different triplets only (Phe, Tyr, His, Gln, Asn, Lys, Asp, Glu, Cys, Arg) suggests C↔T and A↔G mutations to be more frequent. Sueoka quantitative theory of directional mutation pressure indicates that the major cause for a change in DNA GC-content of an organism is the large excess of GC→AT mutations over AT→GC mutations. Also Hildebrand et al. (2010), by examining patterns of synonymous polymorphism using datasets from 149 bacterial species, find a large excess of synonymous GC→AT mutations over AT→GC mutations. This is in accordance also with the fact that if we look at the codon usage (Figure 1) and again we focus our attention on amino acids encoded by two different triplets only (Phe, Tyr, His, Gln, Asn, Lys, Asp, Glu, Cys, Arg), we notice that the more used codons are the GC-poorest ones.

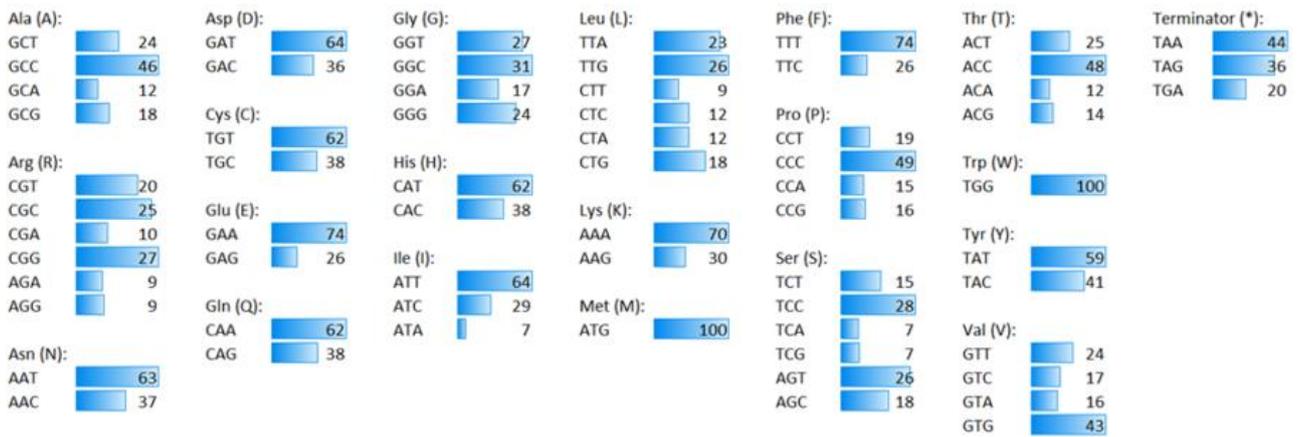

Figure 1. Codon usage. http://2010.igem.org/Design_usu.html

This is probably a defense system evolved to minimize the impact of deleterious mutations originally caused by UV radiations. The UV light first causes two adjacent cytosine residues to form a dimer. During DNA replication, both strands are used as templates to synthesize new strands. The cytosine dimer could cause adenine (instead of the normal guanine) to be incorporated into the new strand. Subsequent DNA replication will produce CC to TT mutation (Figure2).

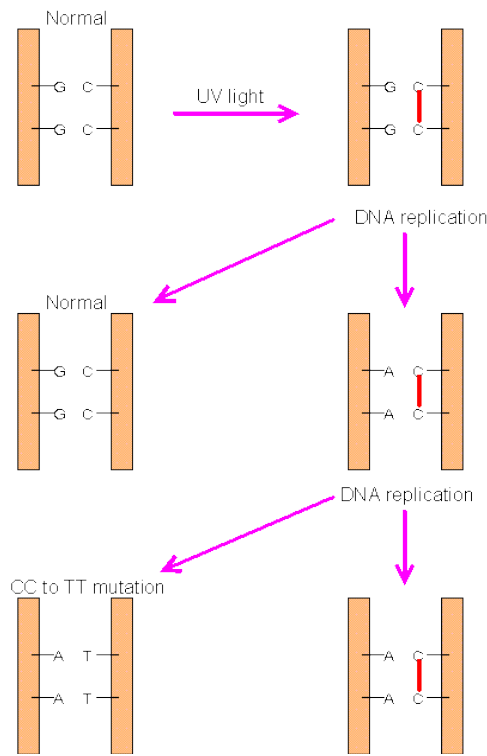

Figure2. G→A and C→T mutations caused by UV radiations.
http://www.web-books.com/MoBio/Free/Ch7F5.htm

All these observations led us to hypothesize mutations of a 3-bonds-forming-nucleotide to a 2-bonds-forming-nucleotide to be not only the more conserved but the more probable too (Agoni, 2013). But actually examining the SNP (single nucleotide polymorphisms) (http://www.ncbi.nlm.nih.gov/snp, http://p53.iarc.fr/SelectedStatistics.aspx) it appears (see Figure3 as an example) that, in addition to GC→AT and GC→TA, AT→GC are within the more frequent mutations.

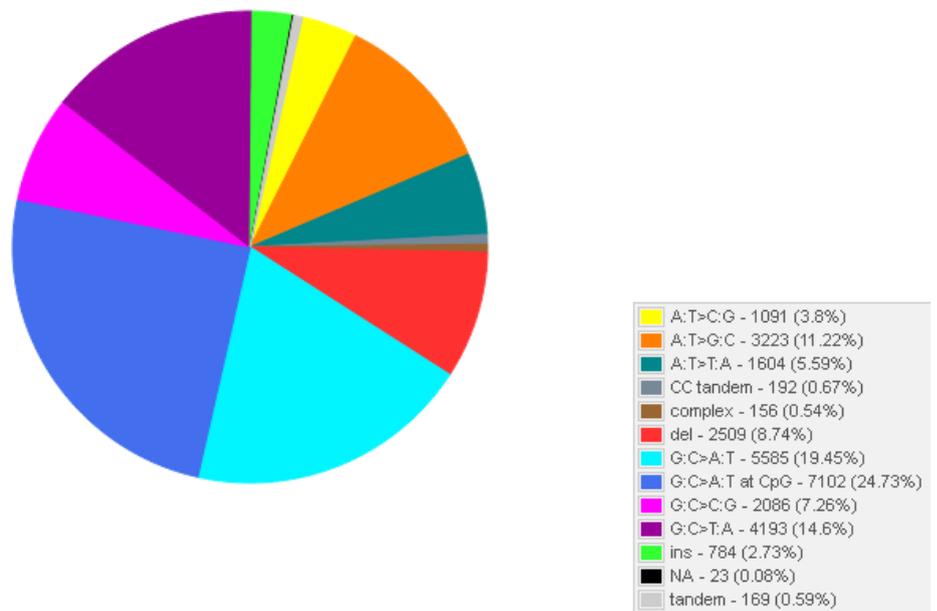

Figure3. Somatic mutations – Mutations pattern (N=28717) IARC TP53 Database, R17, November 2013.

The fact that GC→AT mutations (that group G→A, G→T, C→A, and C→T) are more frequent than AT→GC mutations (A→G plus A→C plus T→G plus T→C) does not mean that this relation is valid for all the subgroups. In other words it is not in contradiction with the observation that T→C can be included into the more frequent mutations overall.

## Fibrillar aggregates and the codon table

Peptides or proteins convert under some conditions from their soluble forms into highly ordered fibrillar aggregates. Such transitions give rise to pathological conditions ranging from neurodegenerative disorders to systemic amyloidoses (Chiti and Dobson, 2006). Not only β-sheet fibrils, but also 4-helical bundles were observed in proteins prone to aggregates formation (Olzscha et al., 2010). Despite recent advances toward the elucidation of the structures of amyloid fibrils and the mechanisms of their formation at a molecular level (Tartaglia et al.,2005; Olzscha et al., 2010), much remains unclear. Tartaglia et al., in 2006 introduced an approach to predict proteins secondary structures based on the knowledge of their amino acid sequences without the inclusion of any additional structural information. These results suggest that the propensity of different regions of the structures of globular proteins to undergo local unfolding events can be predicted from their amino acid sequences with an accuracy of 80% or better (Tartaglia et al., 2005). Their studies revealed that interestingly several five-residue stretches are found in segments that were known to aggregate, this indicating that that can be much important respect to Pα (propensity for α-helices) and Pβ (propensity for β-strands).

At least in principle fibrillar aggregates derive from mutations that create amino acids with higher P$\alpha$ or P$\beta$, but this do not really correspond to the observations.

| Amino Acid | Pα | Pβ |
|---|---|---|
| A | 1.46 (1.42) | 0.78 (0.83) |
| C | 0.75 (0.70) | 1.26 (1.19) |
| D | **0.83 (1.01)** | 0.51 (0.54) |
| E | 1.37 (1.51) | **0.69 (0.37)** |
| F | **0.97 (1.13)** | 1.49 (1.38) |
| G | 0.43 (0.57) | 0.69 (0.75) |
| H | 0.90 (1.00) | **1.05 (0.87)** |
| I | 1.07 (1.08) | 1.65 (1.60) |
| K | 1.14 (1.16) | 0.78 (0.74) |
| L | **1.36 (1.21)** | **1.13 (1.30)** |
| M | **1.25 (1.45)** | 1.10 (1.05) |
| N | 0.72 (0.67) | **0.64 (0.89)** |
| P | **0.40 (0.57)** | **0.32 (0.55)** |
| Q | **1.34 (1.11)** | **0.82 (1.10)** |
| R | **1.24 (0.98)** | 0.88 (0.93) |
| S | 0.76 (0.77) | 0.88 (0.75) |
| T | 0.76 (0.83) | 1.21 (1.19) |
| V | 0.94 (1.06) | **1.89 (1.70)** |
| W | 1.05 (1.08) | 1.39 (1.37) |
| Y | **0.95 (0.69)** | 1.47 (1.47) |

Table 1. Propensity values for α-helices (Pα) and β-strands (Pβ) calculated according to the method of Chou and Fasman (Chou and Fasman, 1978) from a set of 1091 non-redundant proteins. The Pα and Pβ values derived originally from a set of 29 proteins are given in parentheses. P values that differ >0.15 between the two data sets are in boldface type.

The mutations listed in 'Mutations in Hereditary Amyloidosis' database (http://www.amyloidosismutations.com/), known to be associated with diseases characterized by formation of fibrillar aggregates, cover the entire sequence of proteins, not only the five-residue stretches sequences described by Tartaglia.

In addition, it has been recently demonstrated that amyloid-like aggregates sequester numerous metastable proteins with essential cellular functions (Olzscha et al., 2010).

Summarizing fibrillar aggregates can be the result of many unfolded or misfolded proteins of the same type or they can result from the aggregation with many other proteins.

Nevertheless, taking into account Pβ and Pα, the amino acids coding table reveals that unfortunately GC→AT mutations leads to amino acids that have on average a higher pβ. Infact amino acids that are more prone to create β−sheets derive from GC-poorest codons – this considering the third codon as poorely influent - (Figure4). On the other hand Pro, that has the lower Pβ interestingly derives from CC- codons.

|   | C | G | A | T |
|---|---|---|---|---|
| C | CCT Pro P<br>CCC Pro P<br>CCA Pro P<br>CCG Pro P | CGT Arg R<br>CGC Arg R<br>CGA Arg R<br>CGG Arg R | CAT His H<br>CAC His H<br>CAA Gln Q<br>CAG Gln Q | CTT Leu L<br>CTC Leu L<br>CTA Leu L<br>CTG Leu L |
| G | GCT Ala A<br>GCC Ala A<br>GCA Ala A<br>GCG Ala A | GGT Gly G<br>GGC Gly G<br>GGA Gly G<br>GGG Gly G | GAT Asp D<br>GAC Asp D<br>GAA Glu E<br>GAG Glu E | **GTT Val V**<br>**GTC Val V**<br>**GTA Val V**<br>**GTG Val V** |
| A | **ACT Thr T**<br>**ACC Thr T**<br>**ACA Thr T**<br>**ACG Thr T** | AGT Ser S<br>AGC Ser S<br>AGA Arg R<br>AGG Arg R | AAT Asn N<br>AAC Asn N<br>AAA Lys K<br>AAG Lys K | **ATT Ile I**<br>**ATC Ile I**<br>ATA Ile I<br>ATG Met M |
| T | TCT Ser S<br>TCC Ser S<br>TCA Ser S<br>TCG Ser S | **TGT Cys C**<br>**TGC Cys C**<br>TGA stop *<br>**TGG Trp W** | **TAT Tyr Y**<br>**TAC Tyr Y**<br>TAA stop *<br>TAG stop * | **TTT Phe F**<br>**TTC Phe F**<br>TTA Leu L<br>TTG Leu L |

Figure4. Amino acids coding table. Amino acids with high β-strands formation propensity (Pβ) are highlighted in green. They correspond to GC-poorest codons in the amino acids coding table.

Effectively, if we stop here the analysis of mutations related to amyloidosis, they reveal that the majority are GC→AT and lead to an increase in Pβ. Therefore we could speculate that amyloidosis could be a negative side effect of an ancestral defense system against deleterious mutations caused by UV radiations.

But if A:T→G:C mutations are included into the more abundant cathegory (Figure3) we can assert to be in some way protected from amyloidosis by evolution. Infact, a recent paper (Hormoz, 2013) demonstrates that amino acid composition of proteins reduces deleterious impact of mutations.

Balasubramanian et al. have performed a comprehensive analysis of 39,408 SNPs on human chromosomes 21 and 22 from The SNP Consortium (TSC) database, where SNPs are obtained by random sequencing using consistent and uniform methods. They have also performed secondary structure prediction on all coding regions and found that there is no preferential distribution of SNPs in α-helices, β-sheets or coils. This could imply that protein structures, in general, can tolerate a wide degree of substitutions (Balasubramanian).

## Final observations

The codon table protects us from both environment-induced mutations caused by UV radiations or reactive oxygen species (ROS) and from the most spontaneous mutations during replication. Moreover it reflects amino acids abundance (http://www.tiem.utk.edu/~gross/bioed/webmodules/aminoacid.htm).
Maybe that originally the main mutations came from UV or ROS, as a consequence the codon table evolved to minimize their effects. But now the more abundant mutations generated by the polymerase includes A:T→G:C.

Alternatively and more probable, in bacteria, that have an high GC-content, GC→AT mutations are more probable (see Figure5 as an example) respect to organisms with a lower GC-content just for a reason of nucleotide abundance.

Can we hypothesize an evolution of the relative frequencies of the different single nucleotide substitutions to protect humans from amyloidosis?

Can we speculate that amyloidosis can derive from the accumulation of GC→AT mutations with aging? Can amyloidosis be a negative side effect of an ancestral defense system against deleterious mutations caused by UV radiations?

The relationship between mutations and Pβ still remains an open problem due to the difficulty of quantifying the frequency of the diverse nucleotide substitutions.

|   | A | C | G | T |
|---|---|---|---|---|
| A |   | 1 | 12 | 5 |
| C | 8 |   | 2 | 799 |
| G | 561 | 0 |   | 5 |
| T | 1 | 6 | 1 |   |

Figure5. Benomyl pool SNP statistics. (A) Matrix showing the base changes for all new SNPs (N = 1401) in the benomyl pool (Dustin et al., 2014).